\documentclass[aps,pra,twocolumn,final,letterpaper]{revtex4}

\usepackage{graphicx}   
\usepackage{import}                         
\usepackage{epstopdf}
\usepackage{amsmath} 
\usepackage{bm}
\usepackage{amssymb}
\usepackage{quotes}
\usepackage{color}
\usepackage{transparent}
\usepackage{mathtools}
\usepackage{textcomp}
\usepackage{dcolumn}

\usepackage{natbib}

\usepackage{multirow}
\usepackage{cancel} 
\usepackage{mdframed}
\usepackage{color}
\usepackage{bm}
\usepackage{dsfont}
\usepackage{slashed}
\usepackage{soul, color} 
\soulregister\ref{7}  
\soulregister\cite{7} 
\renewcommand{\st}[1]{}

\def\ii{{\mathrm{i}}}
\def\ex{{\mathrm{e}}}

\begin{document}
\rmfamily

\title{Controlling light emission with electron wave interference}
\author{Chitraang Murdia$^{1\dagger}$, Nicholas Rivera$^{1}$, Thomas Christensen$^{1}$, Liang Jie Wong$^{2}$,
John D. Joannopoulos$^{1}$
 Marin Solja{\v{c}}i{\'c}$^{1}$, and Ido Kaminer$^{1,3}$}

\affiliation{$^{1}$ Department of Physics, MIT, Cambridge, MA 02139, USA \\
$^{2}$ Singapore Institute of Manufacturing Technology, Innovis, Singapore 138634, Singapore \\
$^{3}$ Department of Electrical Engineering, Technion $-$ Israel Institute of Technology, Haifa 32000, Israel. \\
$\dagger$ Corresponding author e-mail: murdia@mit.edu}

\noindent	

\begin{abstract}
It is a long standing question whether or not one can change the nature of spontaneous emission by a free electron through shaping the electron wavefunction. On one hand, shaping the electron wavefunction changes the respective charge and current densities of the electron. On the other hand, spontaneous emission of an electron is an incoherent process and can often be insensitive to the shape of the electron wavefunction. In this work, we arrive at an affirmative answer examining Bremsstrahlung radiation by free electron superposition states. We find that the radiation can be markedly different from an incoherent sum of the radiations of the two states because of interference of the radiation amplitudes from the two components of the superposition. The ability to control free electron spontaneous emission via interference may eventually result in a new degree of control over radiation over the entire electromagnetic spectrum in addition to the ability to deterministically introduce quantum behavior into normally classical light emission processes.
\end{abstract}

\maketitle

\noindent

The interaction of free electrons with light is a phenomenon that is ubiquitous in fundamental electrodynamics and also in the development of light sources and accelerator technologies. An important class of these interactions is spontaneous emission by a free electron, which takes many guises such as Cerenkov radiation, transition radiation, Smith-Purcell radiation, Bremsstrahlung, inverse Compton scattering, undulator and synchrotron radiation. These processes have important applications in particle detection, astrophysics, materials characterization, and high-energy light sources \cite{cerenkov,SP,friedman,beam}. 

Despite the great variety of the physics and applications of these emission processes, there are a few common features of these spontaneous emission processes. For example, all of these processes can be well-understood classically \cite{FEL1}. In such a treatment, the electron is described by some time-dependent classical current which radiates into the optical surroundings \cite{jackson}.  As is well known, the expressions for the emitted power derived classically achieve excellent agreement with what is observed barring situations in which the electron recoils heavily from its light emission. What is also fairly well known is that these calculations match a quantum treatment \cite{friedman,gover}. In a quantum treatment, the light emission is captured as a transition between an initial electron state with no photons and a final state with a scattered electron and an emitted photon.  In such a treatment, it must be assumed either that the initial electron is in a single eigenstate or that its density matrix is purely diagonal. This must be the case because electron superposition states are inherently non-classical.

It follows from this discussion that it is interesting to ask: can coherent effects such as electron superposition affect the spontaneous emission? If so, this would be attractive given the recently expanded ability to shape electron wavefunctions into complex spatiotemporal patterns\cite{bliokh,verbeeck,OAM,airy,mask1,mask2,SAD,SQW,feist,FEB,ats}; one could envision designing a partial spacetime profile of the electron wavefunction in order to achieve a target spacetime intensity profile for the emitted light. On the one hand, this question can seem as if the answer is a trivial 'yes' because shaping the electron wavefunction shapes the charge and current density, which classically has a clear effect on radiation. On the other hand, when spontaneous emission happens, it is sensitive to transition current densities and the emission into different final states is incoherent meaning that there are some incoherent aspects of the process. And in fact, there are known cases in which the spontaneous emission power is utterly independent of the initial electron wavefunction. For example, when one considers Cerenkov radiation by an electron superposition, one finds that the long time radiation dynamics are independent of the initial electron wavefunction for reasons of momentum conservation\cite{cerenkovPRX}. Thus, the question still remains: can one shape radiation by shaping the electron wavefunction? In this work, we arrive at an affirmative answer by examining a simple model of Bremsstrahlung radiation by electron superposition states. We find clear evidence for strong interference effects in the emitted radiation.

We start by outlining the model under consideration in Figure 1(a): an electron scattering off of the Coulomb field of a proton fixed in place: $V(x) = -\frac{e^2}{4\pi\epsilon_0 x}$. The initial electron state is taken for simplicity to be an equal probability superposition of two different momenta $\mathbf{p}_1$ and $\mathbf{p}_2$ with relative phase $\xi$:
\begin{equation}
|i\rangle = \frac{|\mathbf{p}_1\rangle + e^{i\xi}|\mathbf{p}_2\rangle}{\sqrt{2}}.
\end{equation}
This state makes a quantum mechanical transition into a final state in which an electron has momentum $\mathbf{r}$ and a photon is emitted with momentum $\mathbf{k}$, the latter of which is parameterized by its frequency $\omega$ and its spherical polar angles $\theta_k$ and $\phi_k$, illustrated in Fig. 1(a). These transitions are characterized by the transition amplitude (or S-matrix element) $S_{fi}$. When squared, multiplied by the photon density of states, and integrated over all electron momenta, it yields the differential power of photon emission by the electron. Since the S-matrix is linear, it takes the form
\begin{equation}
S_{fi} = \frac{1}{\sqrt{2}}\left(S(\mathbf{p}_1\rightarrow\mathbf{r}\mathbf{k}) + e^{i\xi}S(\mathbf{p}_2\rightarrow\mathbf{r}\mathbf{k})\right).
\end{equation}
Interferences will affect the output radiation power if cross terms of the squared S-matrix element do not integrate to zero when integrated over all electron momenta. Importantly, the S-matrices carry delta functions expressing energy-momentum conservation. Generally speaking, provided momentum conservation can be satisfied for two different electron momenta $\mathbf{p}_1$ and $\mathbf{p}_2$ and the same photon momentum $\mathbf{k}$, we expect cross terms to be important. Because the Coulomb field of the scatterer carries a range of momenta, it is possible for different Coulomb-field momenta $\mathbf{q}_1$ and $\mathbf{q}_2$ to provide the necessary momentum to bring $\mathbf{p}_1$ and $\mathbf{p}_2$ to the same photon $\mathbf{k}$.

In the specific case of Bremsstrahlung radiation, it can be shown (see Supplemental Materials (SM)) that the individual S-matrix elements of (2) take the form
\begin{equation}
S(\mathbf{p}\rightarrow\mathbf{r}\mathbf{k}) = (2\pi)^4 \delta^{(4)}(p-r-k-q) \,\, \ii \mathcal{M}(\mathbf{p}\rightarrow\mathbf{r}\mathbf{k})
\end{equation}
with scattering amplitude $\mathcal{M}$ given by 
\begin{multline} 
\ii \mathcal{M}(\mathbf{p}\rightarrow\mathbf{r}\mathbf{k}) =  \frac{-\ii e^3}{\mathbf{q}^2} \bar{u}_{s'} (r)\left[ \slashed{\epsilon}(\mathbf{k}) \frac{(\slashed{r} + \slashed{k}  + m_\ex )}{2 r \cdot k}  \gamma^0 \right. \\
\left.   - \gamma^0  \frac{(\slashed{p} - \slashed{k}  + m_\ex )}{2 p \cdot k} \slashed{\epsilon}(\mathbf{k}) \right] {u}_s (p).   
\end{multline}
where $p$ and $r$ are the initial and final momenta of the electron, $s$ and $s'$ are the initial and final momenta of the electron  $k$ is the photon momentum, and $\epsilon(\mathbf{k})$ is the photon polarization. Also, $e$ and $m_\ex$ are the change and mass of an electron, the $u_s$ are spinors for Dirac fermions, $q$ is the virtual momentum associated with the static Coulomb field, and the $\gamma^{\mu}$ are the Dirac matrices. This notation is summarized in the Supplementary Materials. To complete the discussion of the formalism, we note that the differential cross section per unit photon energy per unit photon solid angle takes the form
\begin{equation}
\frac{\mathrm{d}\sigma}{\mathrm{d}\mathbf{k}} =  \frac{1}{(2\pi)^5} \frac{\omega_\mathbf{k} \beta_\mathbf{r} E_\mathbf{r}}{ 8 \beta_\mathbf{p} E_\mathbf{p} } 
\int \mathrm{d}\Omega_\mathbf{r} 
\left(\frac{1}{2}\sum_{s,s',\epsilon(\mathbf{k})} \left|\ii \mathcal{M}\right|^2\right)
\end{equation}
where $\mathrm{d}\mathbf{k} \equiv \mathrm{d}\omega_\mathbf{k} \mathrm{d}\Omega_\mathbf{k}$, $\omega_\mathbf{k}$ is the photon energy, $E_\mathbf{p}$ and $E_\mathbf{r}$ are the speeds of the initial and final electrons, $\beta_\mathbf{p}$ and $\beta_\mathbf{r}$ are the speeds of the initial and final electrons, and $\Omega_\mathbf{k}$ and $\Omega_\mathbf{r}$ denotes photon and final electron solid angles.

\begin{figure}[t]
\centering
\includegraphics[width=90mm]{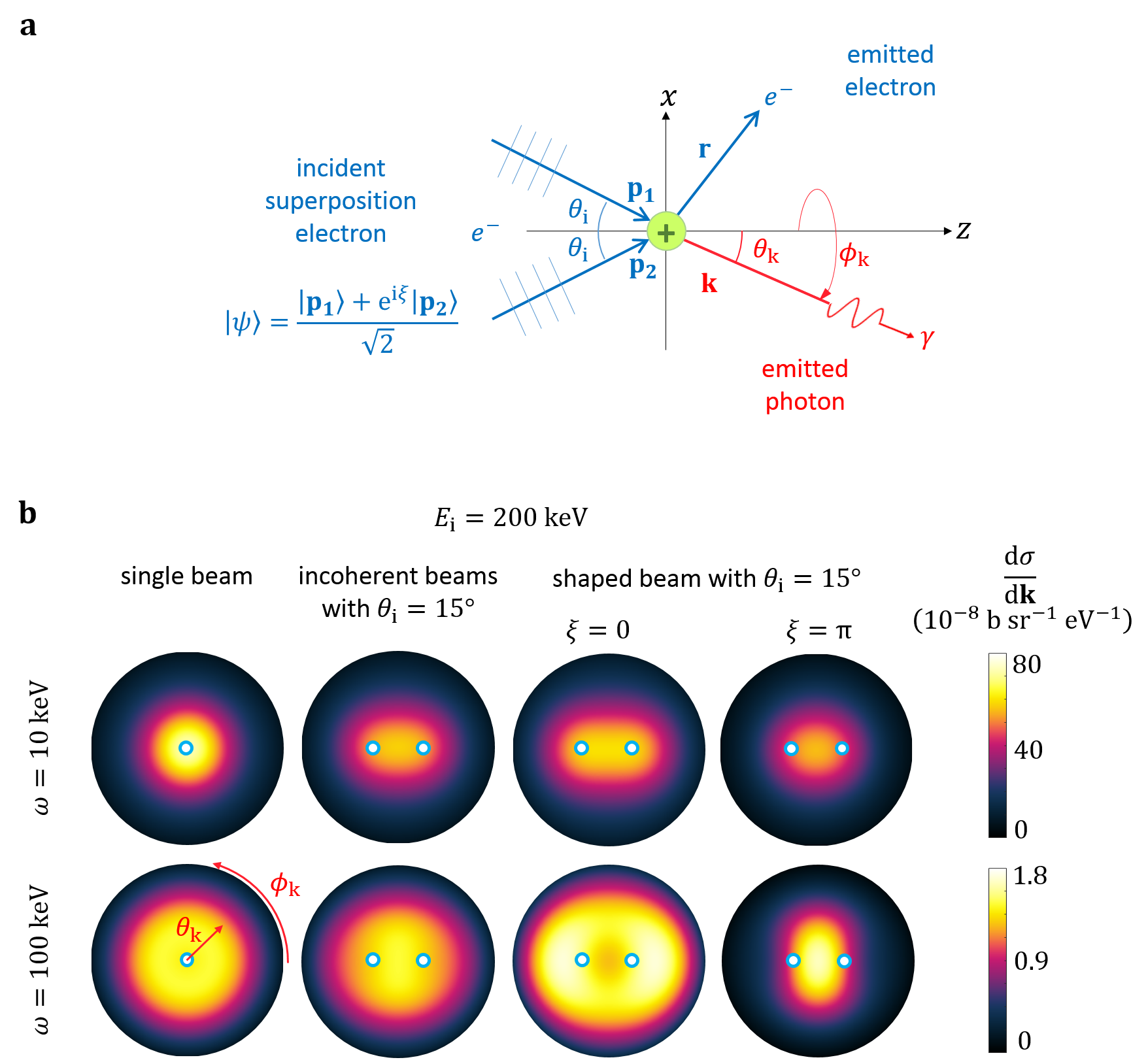}
\caption{\textbf{Shaping radiation with electron interference.} (a) A shaped electron (superposition of two momentum eigenstates) is scattered off a nucleus into a single momentum eigenstate and it emits a Bremsstrahlung photon in the process. (b) Comparison of angular differential cross section (ADCS) for an unshaped electron and a shaped electron. The plots show ADCS in the forward direction and the blue circles indicate the direction of the incident electron momenta. Shaping affects the radiation pattern in a highly non-trivial way.}
\end{figure}

\begin{figure}[t]
\centering
\includegraphics[width=80mm]{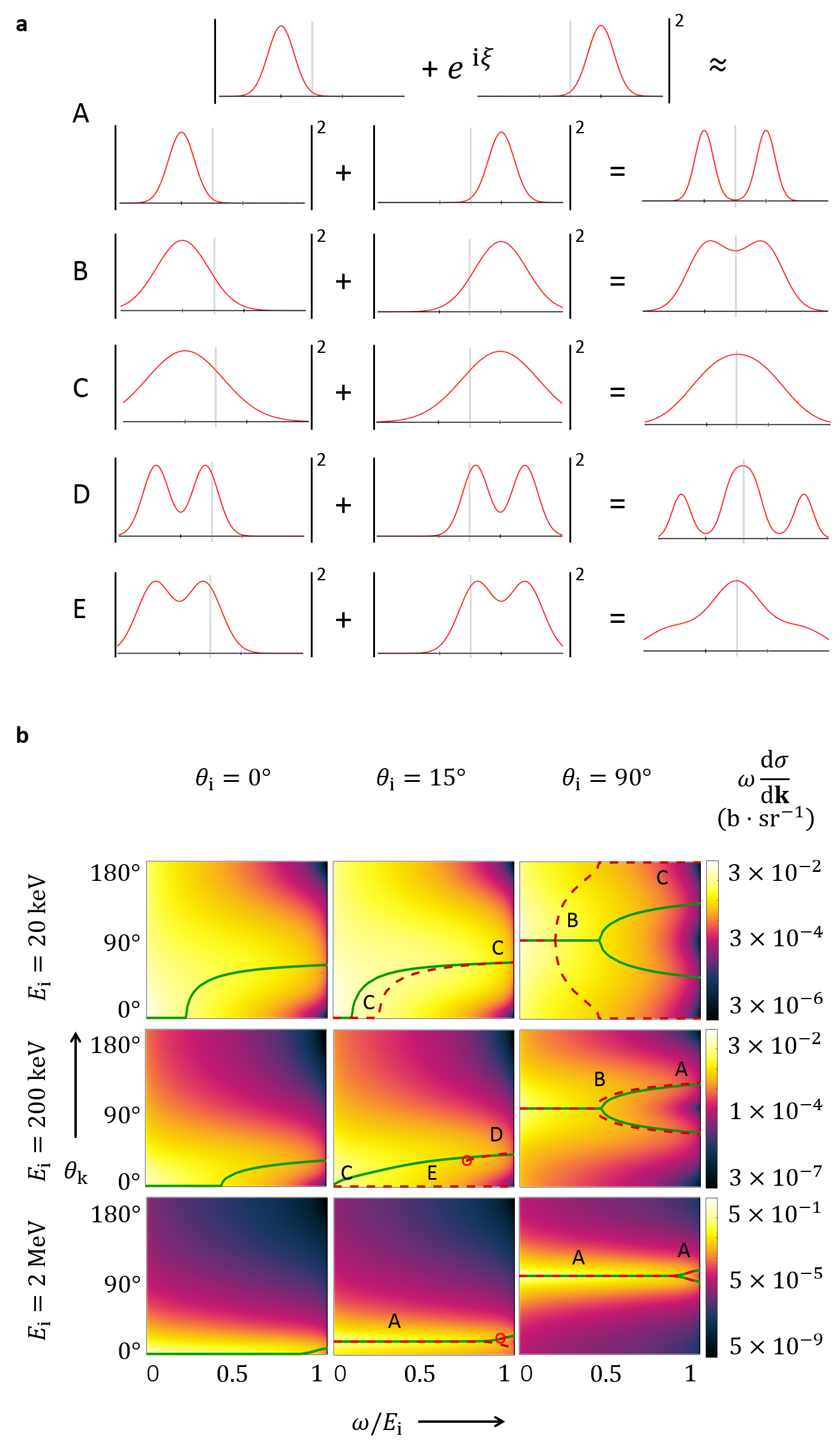}
\caption{ \textbf{Understanding how electron interference affects radiated power.} (a) 'Toy model' to explain the behavior seen in the plots. `A' corresponds to peak width being much smaller than the separation between peaks so no interferences takes place. As the peaks become wider interference effects cause the peaks to shift inwards as in `B'. If the peaks are wide enough, only a single central maxima is seen as shown in `C'. `D' and `E' show analogous plots if there are two peaks in both wavefunctions. (b) A measure of angular differential power (ADP) as a function of detector location  and normalized photon energy. The solid green lines show the direction of maximum ADP at a given photon energy. The dashed red lines show corresponding maxima for an incoherent superposition of two electron beams with the same. The letters `A' through `E' represent the scenarios of our toy model that explain the maxima in the corresponding energy range. Different photon energies show maxima at different location which enables us to use shaped electrons as natural gratings. }
\end{figure}

 We now move to an analysis of the effect of interference on the radiated spectrum. In Figure 1(b), we plot the differential cross section per unit photon energy per unit photon solid angle as a function of the two spherical angles of the photon for two different emission frequencies: 10 and 100 keV. The electron energy is taken to be 200 keV. The blue circles in the figure represent the angles of propagation of the two momenta $\mathbf{p}_1$ and $\mathbf{p}_2$. As can be seen from the figure, the case of the coherent superposition is clearly different from the case of the incoherent superposition for two electron momenta separated by 30\textdegree . In particular, when the two electron states are in phase ($\xi = 0$), the angular width of the radiation is increased relative to the incoherent case, while when they are out of phase ($\xi = \pi$), it is decreased relative to the incoherent case. The decreased width of the angular spectrum of radiation when the electron states are out of phase is an unambiguous signature of destructive interference of amplitudes for radiation at angles at which the spectra associated with $\mathbf{p}_1$ and $\mathbf{p}_2$ overlap. Similarly, an increased peak width can be a sign of constructive interference of amplitudes though it is also possible for the peak widths to increase simply as a result of incoherent summation. In the remaining parts of this work, we explain these results in more detail.


We start by describing the angular spectrum of Bremsstrahlung radiation from a single electron momentum. The spectrum has a strong dependence on electron energy and photon energy. Possible shapes for the transition amplitudes are illustrated in Figure 2(a), columns A-E. Focusing on only the left-most plot of each column signifies the transition elements for a single electron momenta. For high electron energies, it is sharp with angular width $\gamma^{-1} \equiv \sqrt{1-v^2/c^2}$ (case A), while for low electron energies, it is fairly broad in angular width (cases B and C, where C is lower energy). In some cases, the electron spectrum can be double-peaked for higher photon energies. The final thing of note is that the amplitude for radiation can switch sign when the final electron momentum $\mathbf{r}$ becomes antiparallel to the initial electron momentum $\mathbf{p}$. Having explained the individual electron Bremsstrahlung spectra, we are now in a position to describe the coherent and incoherent combinations of these spectra for an electron in a superposition of two momenta. When a second electron is introduced incoherently, we expect the situations on the right-most plots of subcolumns A-E: if the peaks overlap, they can combine to form either two maxima with lower separation due to the increased spectra in between the peaks, or they can form a single maximum because the two peaks are sufficiently close together. In both cases however, the overlap can cause peak shifts even in the incoherent case. The only case where one expects negligible peak shifts is when the electron is highly relativistic such that the angular spread of the radiation is much less than the angle between $\mathbf{p}_1$ and $\mathbf{p}_2$. 

We confirm this intuition in Figure 2(b), where we show in detail the angular and frequency power spectrum of the electron $\omega \frac{d\sigma}{d\mathbf{k}}$. In all cases, the relative phase $\xi$ is taken to be zero. On each panel, the green line represents the peak angle of emission as a function of photon energy. The red dashed line represents that same peak assuming that the two electron states are incoherent with one another. The main phenomenon to see here is that for highly relativistic electrons, the superposition and incoherent cases give nearly identical results, while for lower energy electrons there can be substantial deviations. Given our predictions in the previous paragraph, this is sensible. For the high energy electrons, the two momenta have spectrum with negligible overlap and thus the squared sum of amplitudes is approximately the sum of the squared amplitudes. For lower energies where peak overlap occurs this does not have to be the case. In particular, when there are constructive interferences, the green line move away from the forward emission ($\theta_\mathrm{k} = \theta_\ii$) while for the destructive interferences (due to the sign-flip of amplitudes described in the previous paragraph), the green lines move towards forward emission. For instance, at $E_\ii = 20 \mathrm{keV}$ and $\theta_k = $ 90\textdegree, the green lines are closer to $\theta_\mathrm{k} = $ {90\textdegree} due to destructive interference.

In summary, we have shown that when momentum conservation is relaxed, superposition effects can play a significant role in the emission characteristics of an electron. We have shown this through the specific example of Bremsstrahlung. It is an interesting question to ask if it is possible to see a non-trivial influence of superposition in cases where the electron emission is much more monochromatic, as is the case in Smith-Purcell radiation or undulator radiation or inverse Compton scattering. If it is possible, then the possibility is raised to have new kinds of monochromatic quantum light sources where control over the emission is attained not only through the electromagnetic environment of the electron but also by the spatiotemporal properties of the wavefunction, which can be shaped through masks and pulses. 

\section*{ACKNOWLEDGEMENTS}
We gratefully acknowledge  Nahid Talebi, Claus Ropers, Avi Gover, Roei Remez, Javier Gacia de Abajo, Ady Arie, and Jo Verbeeck for the stimulating and helpful discussions. Research supported as part of the Army Research Office through the Institute for Soldier Nanotechnologies under contract no. W911NF-13-D-0001 (photon management for developing
nuclear-TPV and fuel-TPV mm-scale-systems).Also supported as part of the S3TEC, an Energy Frontier Research Center funded by the US Department of Energy under grant no. DE-SC0001299 (for fundamental photon
transport related to solar TPVs and solar-TEs). N.R. was supported by a US Department of Energy fellowship no. DE-FG02-97ER25308. L.J.W. was supported by the Science and Engineering Research Council (SERC) grant no. 1426500054 of the Agency for Science, Technology and Research (A*STAR), Singapore. I.K. was supported in part by Marie Curie grant no. 328853-MC-BSiCS. 

\section*{SUPPLEMENTARY MATERIAL}

Consider an electron that interacts with the Coulomb potential and emits a photon in the process. Let the initial electron have momentum $p$ and spin $s$, the final electron have momentum $r$ and spin $s'$ and the photon have momentum $k$ and polarization $\epsilon(\textbf{k})$. Let $q$ be the virtual momentum associated with the static Coulomb field. 

We are interested in studying the process $\mathbf{p}\rightarrow\mathbf{r}\mathbf{k}$. The overlap of the incoming and outgoing states is given by 
\begin{equation}
 \sideset{_\mathrm{out}}{_\mathrm{in}} {\mathop{\langle \mathbf{r}\mathbf{k}|\mathbf{p}\rangle}} = \lim_{T \to \infty}\langle \mathbf{r}\mathbf{k}|\ex^{2 \ii H T} | \mathbf{p}\rangle
\end{equation}
where $H$ is the QED Hamiltonian and $T$ is the time interval... . Now, we define the S-matrix element as
\begin{equation}
 \sideset{_\mathrm{out}}{_\mathrm{in}} {\mathop{\langle \mathbf{r}\mathbf{k}|\mathbf{p}\rangle}} \equiv S(\mathbf{p}\rightarrow\mathbf{r}\mathbf{k}).
\end{equation}

The S-matrix element for this process is of the form
\begin{equation}
S(\mathbf{p}\rightarrow\mathbf{r}\mathbf{k}) = (2\pi)^4 \delta^{(4)}(p-r-k-q) \,\, \ii \mathcal{M}(\mathbf{p}\rightarrow\mathbf{r}\mathbf{k})
\end{equation}
where $\mathcal{M}$ is the scattering amplitude.

To leading order, this scattering amplitude is given by 
\begin{multline} \label{M1}
\ii\mathcal{M}(\mathbf{p}\rightarrow\mathbf{r}\mathbf{k}) = (\ii e)^2 \bar{u}_{s'}(r) \left[ \gamma^\mu S_F (r+k)\gamma^\nu  \right.\\   
 \left. + \gamma^\nu  S_F (p-k) \gamma^\mu \right]{u}_s(p)  \,\, {\epsilon_\mu}(\mathbf{k}) (A_c)_\nu (\mathbf{q})
\end{multline}
where $e$ is the charge of the electron, the $u_s$ are spinors for Dirac fermions, the $\gamma_\mu$ are the Dirac matrices, $S_F(.)$ is the fermion propagator, and $(A_c)_\nu$ is the momentum space Coulomb potential.
This formula follows from the Feynman diagrams shown in Figure 3.

Let us now put together the various pieces needed to calculate the scattering amplitude. First, we have the momentum space Coulomb potential given by
\begin{equation}
(A_c)_\nu(\mathbf{q}) = \int d^3 x \,\,  \ex^{\ii \mathbf{q}\cdot\mathbf{x}} (A_c)_\nu(x).
\end{equation}
	
In Coulomb gauge, the Coulomb potential is given by
\begin{equation}
(A_c)_\nu(x) = \left(\frac{e}{4 \pi |\mathbf{x}|},0,0,0\right).
\end{equation}
Thus, the momentum space Coulomb potential is 
\begin{equation}
(A_c)_\nu(\mathbf{q}) = \left(\frac{e}{\mathbf{q}^2},0,0,0\right).
\end{equation}
	
Also, the propagators are
\begin{eqnarray}
S_F (r+k) = \frac{\ii(\slashed{r} + \slashed{k}  + m_\ex )}{(r+k)^2-m_\ex^2} = \frac{\ii(\slashed{r} + \slashed{k}  + m )}{2 r \cdot k}\\
S_F (p-k) = \frac{\ii(\slashed{p} - \slashed{k}  + m_\ex )}{(p-k)^2-m_\ex^2} = - \frac{\ii(\slashed{p} - \slashed{k}  + m )}{2 p \cdot k}
\end{eqnarray}
where $m_\ex$ is the mass of the electron and $\slashed{p} = \gamma^\mu p_\mu$.
	
Substituting these in (\ref{M1}), we get
\begin{multline} \label{M2}
\ii \mathcal{M}(\mathbf{p}\rightarrow\mathbf{r}\mathbf{k}) =  \frac{-\ii e^3}{\mathbf{q}^2} \bar{u}_{s'} (r)\left[ \slashed{\epsilon}(\mathbf{k}) \frac{(\slashed{r} + \slashed{k}  + m_\ex )}{2 r \cdot k}  \gamma^0 \right. \\
\left.   - \gamma^0  \frac{(\slashed{p} - \slashed{k}  + m_\ex )}{2 p \cdot k} \slashed{\epsilon}(\mathbf{k}) \right] {u}_s (p).   
\end{multline}

\begin{figure}[t]
\centering
\includegraphics[width=70mm]{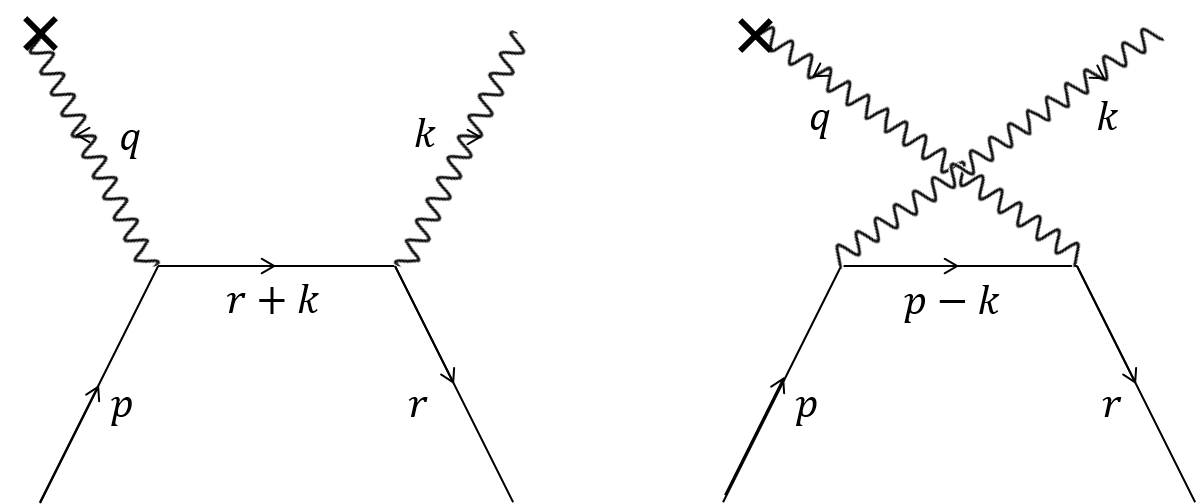}
\caption{Tree-level diagrams for Bremsstrahlung.}
\end{figure}

This matrix element can now be used to calculate the differential cross section. The probability of transition is given by
\begin{equation}
\mathcal{P} = \frac{\left|\sideset{_\mathrm{out}}{_\mathrm{in}} {\mathop{\langle \mathbf{r}\mathbf{k}|\mathbf{p}\rangle}}\right|^2}{\langle \mathbf{p}| \mathbf{p}\rangle\langle \mathbf{r}| \mathbf{r}\rangle\langle \mathbf{k}| \mathbf{k}\rangle}. 
\end{equation}

The normalization factor is given by
\begin{equation}
\langle \mathbf{p}| \mathbf{p}\rangle = (2\pi)^3 \, 2E_\mathbf{p} \delta^{(3)}(0) = 2E_\mathbf{p} V
\end{equation}
where $V$ is the volume...

Thus, the scattering probability is
\begin{align}
\mathcal{P} &= \frac{\left| \langle \mathbf{r}\mathbf{k}|S|\mathbf{p}\rangle\right|^2}{8 E_\mathbf{p} E_\mathbf{r} \omega_\mathbf{k} V^3}\\
&= \frac{(2\pi)^8 \delta^{(4)}(p-r-k-q) \delta^{(4)}(0) }{8 E_\mathbf{p} E_\mathbf{r} \omega_\mathbf{k} V^3} \left|\ii \mathcal{M}\right|^2\\
&= \frac{(2\pi)^4 \delta^{(4)}(p-r-k-q) VT \, }{8 E_\mathbf{p} E_\mathbf{r} \omega_\mathbf{k} V^3} \left|\ii \mathcal{M}\right|^2.
\end{align}

The transition probability per unit time is
\begin{equation}
\mathcal{\dot{P}} = \frac{(2\pi)^4 \delta^{(4)}(p-r-k-q) }{8 E_\mathbf{p} E_\mathbf{r} \omega_\mathbf{k} V^2} \left|\ii \mathcal{M}\right|^2.
\end{equation}

Finally, we have the differential cross section
\begin{equation}
\mathrm{d}\sigma = \dot{P} \left(\frac{V}{(2\pi)^3}\right)^3 \mathrm{d}^3r  \, \mathrm{d}^3k \, \mathrm{d}^3q \, \frac{1}{\beta_\mathbf{p}V}
\end{equation}
where $\beta_\mathbf{p}$ is the speed of the incident electron. Integrating over $q$ using the spatial part of the $\delta$-function,
\begin{equation}
\mathrm{d}\sigma =  \frac{1}{(2\pi)^5} \frac{\delta(E_\mathbf{p} -E_\mathbf{r} -\omega_\mathbf{k})\, \mathrm{d}^3r  \, \mathrm{d}^3k}{8 E_\mathbf{p} E_\mathbf{r} \omega_\mathbf{k} \beta_\mathbf{p}} \left|\ii \mathcal{M}\right|^2
\end{equation}
Note that this fixes $\mathbf{q} = \mathbf{p} - \mathbf{r} - \mathbf{k}$ as required by momentum conservation. Also, We have $q^0 = 0$ by definition.  

Using  polar coordinates,
\begin{multline}
\mathrm{d}\sigma =  \frac{1}{(2\pi)^5} \frac{\delta(E_\mathbf{p} -E_\mathbf{r} -\omega_\mathbf{k}) }{ 8 E_\mathbf{p} E_\mathbf{r} \omega_\mathbf{k} \beta_\mathbf{p}} \times \\
\beta_\mathbf{r} E_\mathbf{r}^2 \mathrm{d}E_\mathbf{r} \mathrm{d}\Omega_\mathbf{r} \, \omega_\mathbf{k}^2 \mathrm{d}\omega_\mathbf{k} \mathrm{d}\Omega_\mathbf{k} \left|\ii \mathcal{M}\right|^2
\end{multline}
where $\beta_\mathbf{r}$ is the speed of the final electron and $d\Omega$ denotes an angular element. 
Integrating over $E_\mathbf{r}$ using the $\delta$-function,

\begin{equation}
\mathrm{d}\sigma =  \frac{1}{(2\pi)^5} \frac{\omega_\mathbf{k}  E_\mathbf{r}\beta_\mathbf{r}}{ 8  E_\mathbf{p} \beta_\mathbf{p}} 
\mathrm{d}\Omega_\mathbf{r} \, \mathrm{d}\omega_\mathbf{k} \mathrm{d}\Omega_\mathbf{k} \left|\ii \mathcal{M}\right|^2
\end{equation}

Till now, we have neglected the electron spins and photon polarizations. To get the total cross section, we average over the initial electron spin, and sum over the final electron spin and photon polarization to get
\begin{multline}
\mathrm{d}\sigma =  \frac{1}{(2\pi)^5} \frac{\omega_\mathbf{k} \beta_\mathbf{r} E_\mathbf{r}}{ 8 \beta_\mathbf{p} E_\mathbf{p} } 
\mathrm{d}\Omega_\mathbf{r} \mathrm{d}\omega_\mathbf{k} \mathrm{d}\Omega_\mathbf{k} \times \\
\left(\frac{1}{2}\sum_{s,s',\epsilon(\mathbf{k})} \left|\ii \mathcal{M}\right|^2\right).
\end{multline}

To obtain the cross section for radiation at a specific angle and frequency, we integrate over the electron solid angle to get the final expression
\begin{equation}
\frac{\mathrm{d}\sigma}{\mathrm{d}\mathbf{k}} =  \frac{1}{(2\pi)^5} \frac{\omega_\mathbf{k} \beta_\mathbf{r} E_\mathbf{r}}{ 8 \beta_\mathbf{p} E_\mathbf{p} } 
\int \mathrm{d}\Omega_\mathbf{r} 
\left(\frac{1}{2}\sum_{s,s',\epsilon(\mathbf{k})} \left|\ii \mathcal{M}\right|^2\right)
\end{equation}
where $\mathrm{d}\mathbf{k} \equiv \mathrm{d}\omega_\mathbf{k} \mathrm{d}\Omega_\mathbf{k}$.

\bibliographystyle{unsrt}
\bibliography{bib.bib}
\end{document}